\documentclass[12pt]{article}

\usepackage[T1]{fontenc}
\usepackage[utf8]{inputenc}
\usepackage{tgpagella}

\usepackage{geometry}               
\usepackage{graphicx}
\usepackage{amssymb}
\usepackage{todonotes}

\usepackage{pdfpages}

\usepackage{eurosym}
\usepackage{amsfonts}
\usepackage{amsmath}
\usepackage{indentfirst}
\usepackage{amssymb}
\usepackage{graphicx}
\usepackage{setspace}
\usepackage{color}
\usepackage[hidelinks]{hyperref}
\usepackage{multirow}
\usepackage[round]{natbib}
\usepackage{rotating}

\usepackage{tikz}
\usetikzlibrary{shapes.geometric, arrows}

\tikzstyle{startstop} = [rectangle, rounded corners, minimum width=3cm, minimum height=1cm,text centered, text width=10cm, draw=black, fill=red!30]

\tikzstyle{process} = [trapezium, trapezium left angle=70, trapezium right angle=110, minimum width=3cm, minimum height=1cm, text centered, text width=10cm, draw=black, fill=blue!30]

\tikzstyle{io} = [rectangle, minimum width=3cm, minimum height=1cm, text centered, text width=10cm, draw=black, fill=orange!30]

\tikzstyle{io2} = [rectangle, minimum width=3cm, minimum height=1cm, text centered, text width=5cm, draw=black, fill=orange!30]

\tikzstyle{decision} = [diamond, minimum width=3cm, minimum height=1cm, text centered, draw=black, fill=green!30]

\tikzstyle{arrow} = [thick,->,>=stealth]

\providecommand{\U}[1]{\protect\rule{.1in}{.1in}}
\providecommand{\U}[1]{\protect\rule{.1in}{.1in}}
\providecommand{\U}[1]{\protect\rule{.1in}{.1in}}
\providecommand{\U}[1]{\protect\rule{.1in}{.1in}}
\providecommand{\U}[1]{\protect\rule{.1in}{.1in}}
\providecommand{\U}[1]{\protect\rule{.1in}{.1in}}
\providecommand{\U}[1]{\protect\rule{.1in}{.1in}}
\providecommand{\U}[1]{\protect\rule{.1in}{.1in}}
\providecommand{\U}[1]{\protect\rule{.1in}{.1in}}

\numberwithin{equation}{section}

\begin{document}

\title{{Cognitive Ability and Tournament Entry: Evidence from Three Korean Populations}}

\author{Syngjoo Choi \footnote{Department of Economics, Seoul National University, 1 Gwanak-ro, Gwanak-gu, Seoul 08826, Republic of Korea. Email: \texttt{syngjooc@snu.ac.kr}}\hspace{8mm} Byung-Yeon Kim \footnote{Department of Economics, Seoul National University, 1 Gwanak-ro, Gwanak-gu, Seoul 08826, Republic of Korea. Email: \texttt{kimby@snu.ac.kr}} \hspace{8mm} Jungmin Lee \footnote{Department of Economics, Seoul National University, 1 Gwanak-ro, Gwanak-gu, Seoul 08826, Republic of Korea. Email: \texttt{jmlee90@snu.ac.kr}} \hspace{8mm} Sokbae Lee \footnote{Corresponding author. Department of Economics, Columbia University, 420 West 118th Street, New York, NY 10027, USA; and Institute for Fiscal Studies, 7 Ridgmount Street, London WC1E 7AE, United Kingdom. Email: \texttt{sl3841@columbia.edu}} \hspace{8mm}}

\date{August 2026}
\maketitle


\begin{abstract}
We compare tournament entry among three Korean groups raised in different institutional environments: South Korea, North Korea, and China. Experiments with a random bonus show that North Korean (NK) refugees enter tournaments less often than South Korean (SK) participants and Korean-Chinese immigrants. The conditional NK-SK difference becomes statistically insignificant when Raven score is included. A choice model with probability weighting suggests that lower cognitive ability is associated with lower expected performance, more pessimistic beliefs, and, within the model, greater aversion to competition.

\vspace{2ex}


\vspace{2ex}

\noindent \textbf{Keywords}: Piece Rate; Tournament; North Korea; Institution; Laboratory Experiment.
\end{abstract}

\newpage




\section{Introduction}

Tournament entry is widely used to measure willingness to compete. Yet the decision to enter a tournament also depends on expected performance, beliefs about winning, risk preferences, and probability weighting. Differences in entry across populations can therefore reflect several dimensions of choice rather than a single competitive preference. We examine this issue among three Korean participant groups and ask how measured cognitive ability relates to tournament entry and its component parts. 

We compare three groups now living in South Korea: participants born and raised in South Korea (SK), North Korean refugees born and raised in North Korea (NK), and Korean-Chinese immigrants born and raised in China (KC). The three groups share Korean ethnicity, historical origins, and a common language, but they grew up under sharply different economic and political institutions. South Korea has a market economy and democracy, whereas North Korea has a centrally planned economy and dictatorship. China combines one-party communist rule with a market economy.
Their shared background and different upbringings provide an unusual setting in which to examine how cross-population entry gaps align with measured components of choice.

To measure tournament entry, we adapt the design of \citet{Niederle:Lise:07}, further used by \citet{Bartling:et:al:09} and \citet{Dohmen:Falk:11} among others, and introduce a random bonus. This bonus provides exogenous individual-level variation in the task score used to determine earnings and tournament outcomes.
 Because performance differs across the three Korean groups, the bonus expands the support of the performance-plus-bonus distribution, particularly for North Korean subjects,\footnote{See \citet{CKLL} for evidence of an NK-SK Raven-score gap.} and enables lower-performing participants to enter the tournament. It also provides variation that we use in the choice model to distinguish components of tournament entry.

The experiment also measures subjective winning probability, risk aversion, and cognitive ability. After completing three rounds of the real-effort task, participants report their beliefs about the probability of winning the tournament, elicited using the binarized scoring rule of \citet{Hossain:Okui:13}.
We elicit risk preferences using the multiple price list design of \citet{Holt:Laury:02} and measure cognitive ability using an abbreviated version of the Raven's Progressive Matrices Test.\footnote{For economic applications of Raven scores, see \citet{Burks:09, Charness:11, Gill:Prowse:16}.}
The Raven test is especially useful in this setting because the three groups have markedly different schooling and life experiences, while the test is a widely used measure of analytic intelligence.\footnote{In psychology, analytic intelligence is defined as ``the ability to reason and solve problems
involving new information without relying extensively on an explicit base of
declarative knowledge derived from either schooling or previous experience''
\citep{Carpenter:90, Nisbett:12}.}

North Korean refugees are significantly less likely than South Korean natives to enter the tournament. The unconditional difference is about 20 percentage points and remains about 10 percentage points after controlling for risk aversion and task performance measured before the compensation choice. North Korean participants also report significantly lower subjective probabilities of winning. By contrast, we find no statistically significant differences between Korean-Chinese immigrants and South Korean natives in either tournament entry or subjective winning beliefs.

Adding the Raven score to the tournament-entry regression changes the sign of the NK-SK coefficient and renders it statistically insignificant. Thus, conditional on measured cognitive ability and the other controls, the estimate provides no evidence of a difference in tournament entry between the two groups. In the post-experiment survey, however, North Korean participants report relatively favorable attitudes toward everyday competition. These survey responses complement the incentivized choice data by describing stated attitudes toward competition.

Nor do we find statistically significant associations between tournament entry and the measured NK-specific experiences, including educational attainment in North Korea and Communist Party membership, a proxy for socioeconomic status in North Korea.
Following recent discussions about variation in willingness to compete \citep{gillen2019experimenting, Veldhuizen}, we use a choice model with probability weighting\footnote{\citet{CKLL2022} study cognitive ability and probability weighting among North Korean refugees and South Koreans using a separate sample. In the present paper, probability weighting is one component of a model of tournament entry.} to relate the Raven score to three components of tournament entry: expected performance, subjective winning beliefs, and, within the model, aversion to competition.
 The random bonus serves as an instrument for subjective winning probability to address measurement error. The model provides a common accounting framework for interpreting the reduced-form group differences and the Raven-score associations. Within this framework, lower cognitive ability is associated with lower expected performance, more pessimistic beliefs, and greater aversion to competition. Together, these estimates show how cognitive ability is related to several components of tournament entry rather than to a single competitive preference.

The comparison has several limitations. The three groups differ in observed and unobserved ways beyond their institutional backgrounds, so the observed differences cannot be attributed solely to institutional experience. North Korean refugees are also a selected and unrepresentative sample of the population living in North Korea, and selection may contribute to both their tournament choices and their Raven scores.\footnote{See \citet{KCLLC}.} Mistrust of authorities or other participants could also influence tournament entry if refugees doubt the integrity of the competition.

More broadly, the design does not isolate the causal effects of the environments associated with the division of Korea. North Korea combines communism and dictatorship, and the data do not separately identify the roles of poverty, malnutrition, or education quality. The evidence therefore provides a case study of tournament entry across three Korean groups rather than a causal decomposition of their institutional environments. 
North Korea nevertheless offers a distinctive setting: its current regime has lasted more than 70 years and has kept its citizens largely isolated from the rest of the world. This history also limits direct comparisons with evidence from other settings.
We postpone discussion of the literature to Section \ref{sec:literature}.

The remainder of the paper proceeds as follows. 
In Section \ref{sec:background}, we briefly review distinct characteristics of three groups of Koreans. 
In Section \ref{sec:sampling}, we describe the sampling and experimental design of our study. 
In Section \ref{sec:results}, we first provide the summary statistics of baseline variables and Raven test scores across the three groups and then present experimental results.
In Section \ref{sec:structural:est}, we develop a choice model between the piece-rate and
 tournament schemes and use it to organize associations between the Raven score and components of tournament entry.
In Section \ref{sec:literature}, we relate our study to several strands of the literature.
Section \ref{sec:conclusion} concludes.
The online appendix includes detailed experimental instructions in Korean and English.

\section{Background}\label{sec:background}

The three groups share a language and historical origins but differ in their institutional and migration backgrounds. This section summarizes the features most relevant to the comparison.

\bigskip

\textbf{North Korean Refugees}\hspace*{1ex}
South and North Korean participants grew up under different economic and political institutions: capitalism and democracy in the South and communism and dictatorship in the North. These systems were installed in the late 1940s and further entrenched after the Korean War
\citep{AJR:2005, KCLLC}. 

Socialist institutions emphasize state ownership and cooperation, and their theorists viewed capitalist competition as a source of anarchy and self-destruction \citep{Marx, Hilferding}. Limited exposure to markets and anti-competition ideology may therefore be associated with lower tournament entry among North Korean refugees. Communist governments nevertheless used state-directed competition to pursue production goals, including the Soviet Stakhanovite movement and North Korea's Chullima movement and speed battles \citep{Prokhorov, Kim2017}. These collective practices differ from individual tournament entry but show that competition can coexist with communist institutions. Whether North Korean refugees enter tournaments less often is ultimately an empirical question.

North Korean refugees are a selected sample: relative to the North Korean population, they are more likely to be female, from bordering provinces, and unmarried \citep[][Table 2]{KCLLC}. North and South Korean participants also differ in other observed and unobserved dimensions, so their tournament-entry differences cannot be attributed solely to institutional experience. To improve comparability, we oversampled low-income South Koreans and controlled for a rich set of demographic and socioeconomic characteristics.

\bigskip

\textbf{Korean-Chinese Immigrants}\hspace*{1ex}
Korean-Chinese immigrants are descendants of Koreans who emigrated to China from the late 19th through the first half of the 20th century. Most grew up and were educated in China before moving to South Korea, mainly for economic opportunities. Their institutional background can be characterized as politically centralized but economically decentralized \citep{Xu:11}. Like North Korean refugees, they are immigrants in South Korea and have similar accents and appearances; unlike the refugees, they grew up with greater exposure to a market economy. If home-country economic institutions are associated with tournament entry, Korean-Chinese participants may resemble South Korean participants.

\bigskip

\section{Sampling and Study Design}\label{sec:sampling}

\subsection{Sampling}

Our study involved three distinct groups of Korean people in terms of their
countries of origin. We used the stratified sampling method to recruit 191
North Korean refugees who were representative of the population of North
Korean refugees in South Korea in terms of gender, age (at least 20 years
old), and year of entry.\footnote{According to the 2014 official statistics of the Ministry of Unification in
South Korea, the population of North Korean refugees consisted of about 28\%
in their twenties, 30\% in their thirties, 16\% in their forties, and 10\%
in their fifties or above. About 28\% of them entered South Korea prior
to 2005, 27\% between 2006 and 2008, 29\% between 2009 and 2011, and the
rest since 2012.}
 In addition, we recruited 193 South Korean adults and 72 Korean-Chinese participants in
South Korea to match  the composition of North Korean
refugee subjects with regard to gender and age. To reduce income
differences between South Korean and North Korean subjects, we oversampled
low-income South Korean subjects by recruiting one-third of the South Korean sample from households
with monthly income below about USD 2,200.

In recruiting these three subject groups
and implementing our experiments, we collaborated with a branch of a global
survey company, the Nielsen Company in Korea, which had accumulated
experience in conducting surveys with representative samples of South
Korean adults and North Korean refugees in South Korea.

\subsection{Study Design}

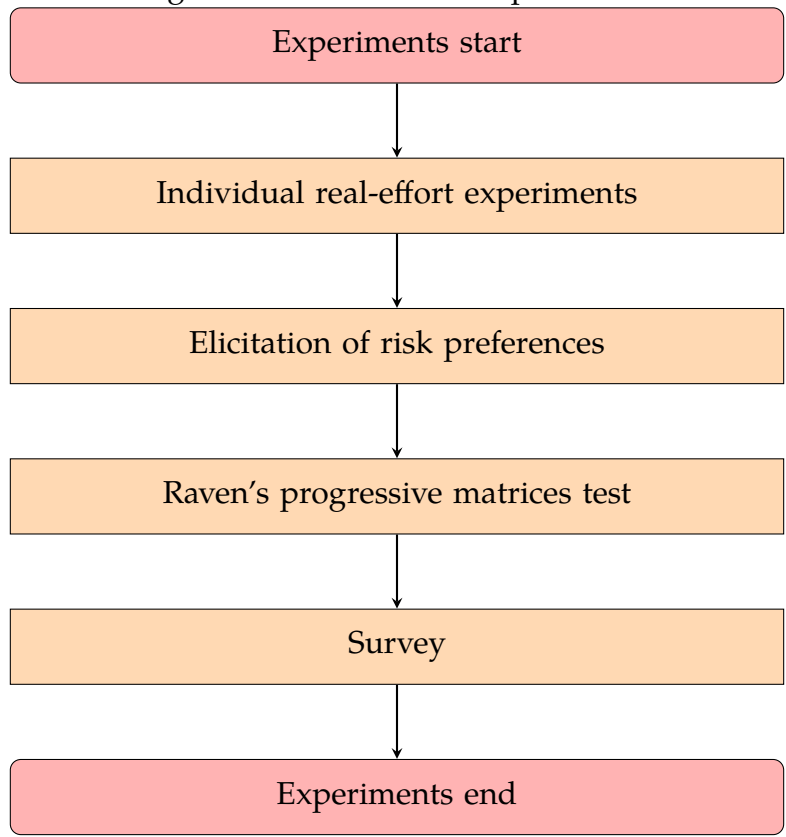
\begin{figure}[htp]
\begin{center}
\caption{Flow Chart of Experiments}
\label{fig-fce}
\begin{tikzpicture}[node distance=2cm]
\node (start) [startstop] {Experiments start};
\node (in1) [io, below of=start]  {Individual real-effort experiments};
\node (in2) [io, below of=in1]  {Elicitation of risk preferences};
\node (in3) [io, below of=in2]  {Raven's progressive matrices test};
\node (in4) [io, below of=in3]  {Survey};
\node (end) [startstop, below of=in4] {Experiments end};

\draw [arrow] (start) -- (in1);
\draw [arrow] (in1) -- (in2);
\draw [arrow] (in2) -- (in3);
\draw [arrow] (in3) -- (in4);
\draw [arrow] (in4) -- (end);

\end{tikzpicture}
\end{center}
\end{figure}

We ran 12 experimental sessions in June 2015 at the Nielsen Company in Korea.
The composition of each session approximately reflected the overall sample: the numbers of North Korean refugees and native-born South Koreans were similar and exceeded the number of Korean-Chinese immigrants.
 Our study design is broadly depicted in  Figure \ref{fig-fce}.

\subsubsection{Individual Real-Effort Experiments}

\textbf{Individual Real-Effort Tasks} \hspace*{1ex}
At the first
stage, subjects conducted a series of individual real-effort tasks under two
different payment schemes. 
 The task in each scheme involved counting $0$'s in a $7\times 7$ table
containing $0$'s and $1$'s \citep[see, e.g.,][]{Abeler2011}.
 Specifically, subjects received 20 of these tables in an envelope,
 counted $0$'s, and entered their answers in a computer within 5 minutes.


Subjects performed this task  under a noncompetitive
piece-rate incentive scheme as well as under a competitive tournament scheme. In the
piece-rate incentive scheme, the subject was paid 1,000 KRW (about one USD) for each correct table.
For example, the subject was paid 12,000 KRW if the answers for 12 tables were correct.
Under the tournament incentive scheme, subjects were informed that each
person would be randomly matched with an anonymous partner at the end of the
session and would earn 2,000 KRW for each correct table if the number of correct
tables made by the subject was higher than that by his/her partner or if the
subject was randomly selected in case of a tie.
 Otherwise, the subject received nothing. 
For example, suppose that subject A answered 9 tables correctly. 
If partner B had 8 correct tables, then A earned 18,000 KRW and B nothing;
if B had 10 correct tables, then A earned nothing and B  20,000 KRW;
if B had 9 correct tables, then a random winner (A or B) earned  18,000 KRW.
 
 We randomized the
order of the two schemes among subjects. At the end of the task under each
payment scheme, individual subjects were informed of the number of correct
answers they made. Under the tournament scheme, whether the individual won
or not was  revealed after they finished all tasks, including the post-experimental survey.

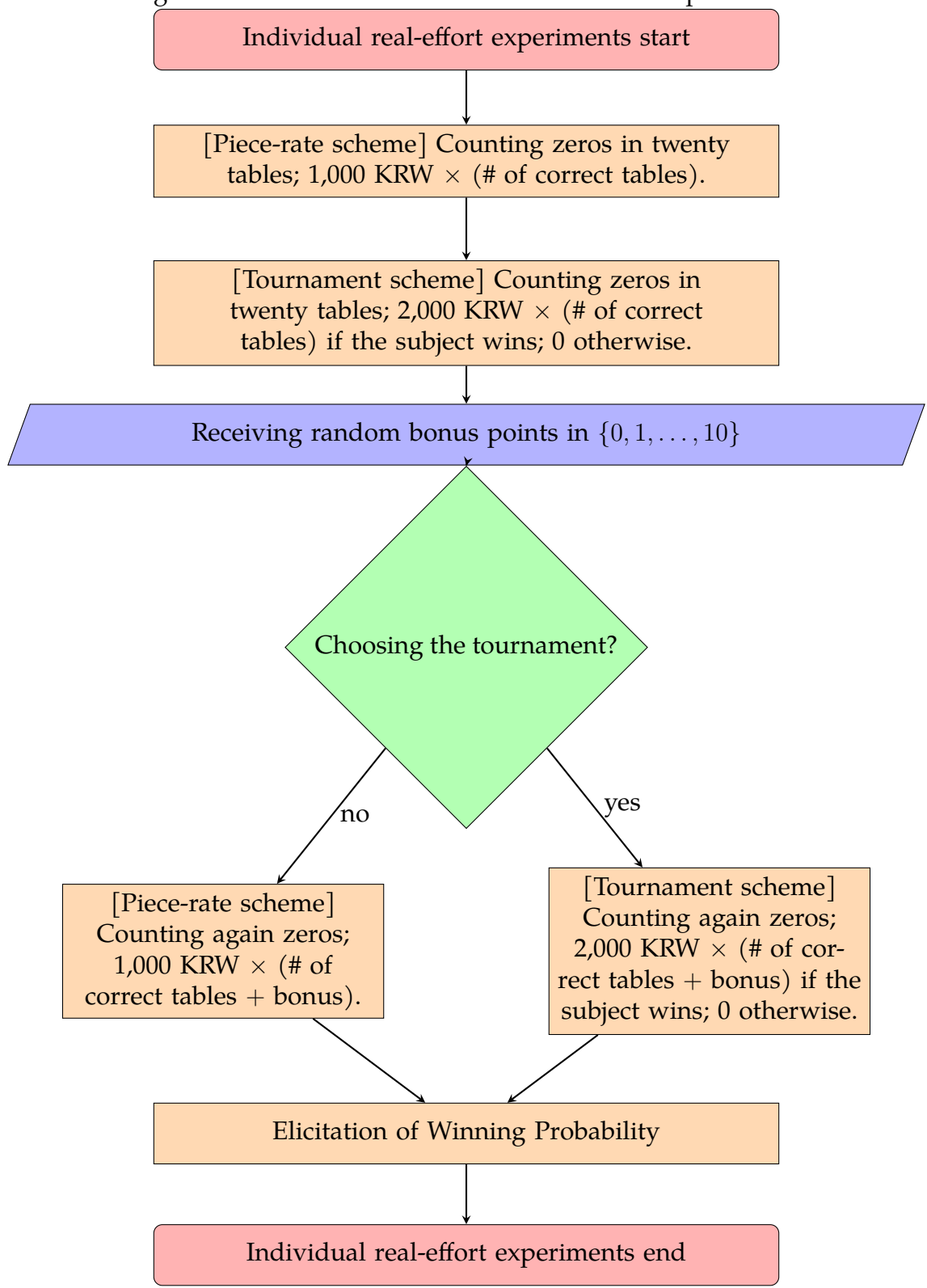
\begin{figure}[htp]
\begin{center}
\caption{Flow Chart of Individual Real-Effort Experiments}
\label{fig-fc-individual}
\begin{tikzpicture}
\node (start) [startstop]  at (0,10) {Individual real-effort experiments start};
\node (in1) [io]  at (0,8)  {[Piece-rate scheme] Counting zeros in twenty  tables; 1,000 KRW $\times$ (\# of correct tables).};
\node (in2) [io]  at (0,5.5)  {[Tournament scheme] Counting zeros in twenty  tables; 2,000 KRW $\times$ (\# of correct tables) if the subject wins; 0 otherwise.};
\node (bonus) [process] at (0,3.5) {Receiving random bonus points in $\{0,1,\ldots,10\}$};
\node (middle) [decision] at (0,0) {Choosing the tournament?};
\node (in3) [io2]  at (-4,-5)  {[Piece-rate scheme] Counting again zeros; 
1,000 KRW $\times$ (\# of correct tables $+$ bonus).};
\node (in4) [io2]  at (4,-5)  {[Tournament scheme] Counting again zeros; 
2,000 KRW $\times$ (\# of correct tables $+$ bonus) if the subject wins; 0 otherwise.};
\node (in5) [io]  at (0,-8)  {Elicitation of Winning Probability};
\node (end) [startstop]  at (0,-10) {Individual real-effort experiments end};

\draw [arrow] (start) -- (in1);
\draw [arrow] (in1) -- (in2);
\draw [arrow] (in2) -- (bonus);
\draw [arrow] (bonus) -- (middle);
\draw [arrow] (middle) -- node[anchor=west] {no} (in3);
\draw [arrow] (middle) -- node[anchor=west] {yes}  (in4);
\draw [arrow] (in3) -- (in5);
\draw [arrow] (in4) -- (in5);
\draw [arrow] (in5) -- (end);
\end{tikzpicture}
\end{center}
\end{figure}

\bigskip

\textbf{Choice of Incentive Schemes} \hspace*{1ex}
After completing the real-effort tasks under the two
incentive schemes, subjects were asked to choose an
incentive scheme under which they performed the same task of counting $0$'s
with a new set of 20 such tables as in the first two rounds. Since there might
have been performance differences among three Korean groups, we introduced an
exogenous variation in the task performance by randomly giving an individual
an integer bonus point between 0 and 10. If the piece-rate scheme was
selected, the individual obtained total earnings as the sum of the number of
correct answers and the bonus point, multiplied by 1,000 KRW. If the
tournament was selected, the individual received the total earnings of the sum of
the number of correct answers and the bonus point, multiplied by 2,000 KRW, if
the individual won and  nothing otherwise. Under the tournament
incentive scheme, the individual with a bonus point competed with the
opponent matched in the previous round under tournament. In deciding the winner, the opponent 
did not have a bonus point. Subjects were informed of this information.

One of the main research questions is why tournament-entry rates differ across the three groups.
A participant may choose the piece-rate scheme because of aversion to competition or lower expected performance relative to an opponent.
The random allocation of bonus points helps distinguish the contribution of performance from other components of tournament entry within the choice model.

\bigskip

\textbf{Elicitation of Winning Probability} \hspace*{1ex}
Upon completing the three rounds of the real-effort experiment, we elicited
subjects' beliefs about their winning probability
using the method of \citet{Hossain:Okui:13}.  Subjects
were reminded of the number of correct answers under the incentive scheme
they chose as well as the bonus point assigned to themselves, and that their
opponent in the stage of belief elicitation was another participant who was
matched under the tournament scheme and had no bonus point. In the case
where the piece-rate scheme was chosen, the subject was asked for his or her
beliefs of the winning probability if he or she had chosen the tournament instead.
Subjects were asked to choose beliefs of their winning probability in the
range between 0\% and 100\% with 10\% increments. We computed a prediction
error based on whether they won and their elicited beliefs. The computed
prediction error was compared with a random number generated between 0 and
1. If the prediction error was lower than a random number, the subject
received 2,000 KRW and  nothing otherwise.
Figure \ref{fig-fc-individual} presents the flow chart of individual real-effort experiments.\footnote{After elicitation of winning probability, participants also completed team-level real-effort tasks in randomly formed groups of three under an exogenously assigned payment scheme. These tasks are not analyzed here because they are not directly related to the paper's research question.}

\subsubsection{Other Measurements}

\textbf{Elicitation of Risk Preferences} \hspace*{1ex}
After team-level experiments,  we elicited
subjects' risk preferences. Specifically,  
we used the multiple price list design of \citet{Holt:Laury:02} for the
elicitation of risk preferences. The design involves ten choices between the
paired lotteries. 
One lottery in the
pair (`safe' choice)  involves 5,000 KRW with probability $p$ and 4,000 KRW with probability $%
1-p$, whereas the other lottery (`risky' choice) entails 10,000 KRW with probability $p$ and
0 KRW with probability $1-p$. The probability of the high-payoff outcome begins with $p=0.1$ and 
increases by $0.1$ as the decision goes down along the list.
The expected payoff difference between the safe and risky lotteries is $4 - 9p$ in 1,000 KRW;
therefore, as a benchmark, a risk-neutral subject's optimal behavior is to make the first four safe choices and then to switch to six risky lotteries. 

\bigskip

\textbf{Raven's Progressive Matrices Test} \hspace*{1ex}
At the end of the experiments, we measured the cognitive ability of
subjects using an abbreviated version of the Raven's Progressive Matrices Test.
Subjects solved 24 questions within 15 minutes. 
 We did not provide any monetary
incentives for completing the Raven test, which is conventional in the
psychology and psychometric literature \citep{Gill:Prowse:16}.

\bigskip

\textbf{Survey} \hspace*{1ex}
At the end of each session, the participants completed a survey that collected information on their demographic and socioeconomic characteristics. We asked several additional questions of North Korean refugees and Korean-Chinese participants.

\section{Empirical Results}\label{sec:results}

In this section, we first provide the summary statistics of baseline variables and Raven test scores across the three groups and then present experimental results.

\subsection{Summary Statistics of Baseline Variables}

\begin{table}[htbp]
\caption{\label{Tab_data_01} Summary Statistics}\centering\medskip
\begin{tabular}{lcccr} \hline \hline
 & NK  & SK  & KC  & p-value  \\  \hline 
Female$\ast$ &       .66 &       .71 &      .708 &      .533 \\  
Age &      37.5 &      34.8 &      33.6 &    .00376 \\  
Post-secondary education$\ast$ &      .262 &       .85 &      .597 &   $<$.001  \\  
Married$\ast$ &      .298 &      .425 &      .306 &     .0246 \\  
Subject health status: not healthy$\ast$ &      .304 &       .14 &      .125 &   $<$.001 \\  
Religious affiliation$\ast$ &      .597 &      .585 &      .417 &      .023 \\  
Number of household members &      2.38 &      3.15 &       2.9 &  $<$.001 \\  
Employed$\ast$ &      .639 &      .798 &      .792 &    .00126 \\  
Unemployed$\ast$ &      .136 &     .0777 &      .111 &      .174 \\  
Out of labor force$\ast$ &      .225 &      .124 &     .0972 &    .00884 \\  
Stock market participation$\ast$ &     .0838 &      .275 &      .139 &  $<$.001 \\  
Credit card holding$\ast$ &      .382 &      .705 &      .333 &  $<$.001 \\  
Online shopping$\ast$ &      .429 &      .933 &      .611 &  $<$.001 \\  
Monthly household income &       1.5 &      4.42 &      2.12 &  $<$.001 \\  
Monthly household income per person &      .794 &      1.64 &      1.01 &  $<$.001 \\  
Monthly household expenditure &      1.05 &      3.34 &      1.54 &  $<$.001 \\  
Household wealth &       133 &       352 &       185 &     .0577 \\  
\hline \end{tabular}
\medskip \\
\parbox{5.5in}{Notes: The table shows the mean  of each variable for  North Korean refugees (NK),
native-born South Koreans (SK) and Korean-Chinese immigrants (KC) separately. The last column shows the p-value for testing the joint significance of group indicators in regressing each variable on group indicators. 
The omitted group is SK in the regression.
Household income, expenditure and wealth are measured in 1 million KRW. The variables with $\ast$ are binary indicator variables. The labor force status consists of employed, unemployed and out of labor force.}
\end{table}

The basic sociodemographic characteristics of the three subject groups are
reported in Table \ref{Tab_data_01}, disaggregated by country of origin (SK, NK, and KC).
A significant majority of each group are female: 66\%
of the NK subjects and about 71\% of each of the SK and KC
subjects. The NK subjects are on average 38 years old, while
the SK subjects and the KC subjects are 35 years old and 34
years old, respectively. 
	
The three Korean groups are different on many observables. 
Less than 30 percent of the NK subjects were post-secondary educated in contrast to 85 percent for the SK subjects and 60 percent of the KC subjects.    Compared to SK, the NK and KC subjects were less likely to be married. 
About 30 percent of NK subjects reported being unhealthy, a much higher share than among SK and KC subjects.
The KC subjects are noticeably less affiliated with religion than both NK and SK. 
The NK subjects had  fewer household members. 
In terms of labor force status, NK participants were least likely to be employed, while there was no difference between SK and KC participants.
Relative to SK, both NK and KC were less likely to participate in the  stock market, to hold a credit card and to shop online. In short, as immigrants, 
both NK and KC were less engaged with various economic aspects than the native SK.
Although we oversampled lower-income SK, their average household income, expenditure and wealth were significantly higher than those of NK and KC. 	

\subsection{Raven Test Results}

\begin{figure}[htbp]
\caption{\label{Fig_Raven} Raven Test Z-Scores}\centering
\medskip
 A. Histogram of Test Scores by Group

\includegraphics[scale=0.9]{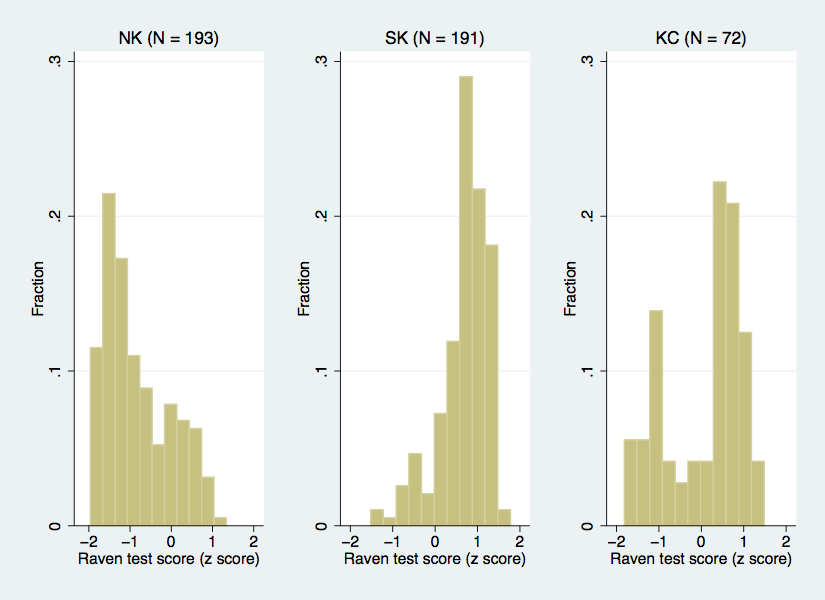} 

\medskip
 B. Age Profile of Test Scores by Group
 
 \includegraphics[scale=0.9]{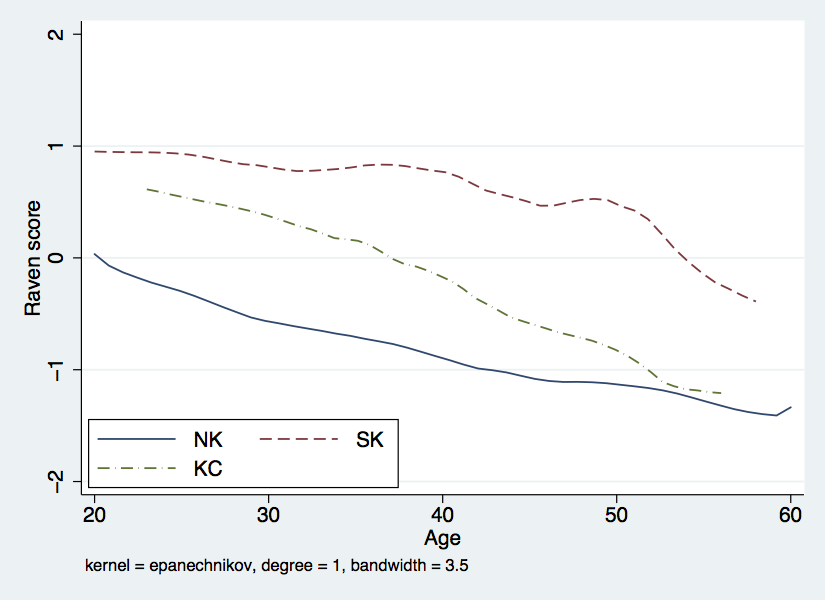} 

\end{figure}

All subjects took the Raven test at the end of experiments. 
Test results are presented in Table \ref{Tab_Raven} and Figure \ref{Fig_Raven}. 
Specifically, 
the summary statistics are given in Table \ref{Tab_Raven};
Panel A of Figure \ref{Fig_Raven} plots the histograms of the test scores
and Panel B shows the age profiles that are obtained by local linear estimates. 
The results reveal a substantial gap between the NK and SK participants in terms of cognitive ability. The average z-score of the SK participants is
0.724, whereas the average score of the NK subjects is only -0.771. The average score of the KC subjects is between those of SK and NK and is slightly above zero.
The Raven test results for SK and NK are quantitatively similar to those reported in our previous work \citep{CKLL}, where Raven test performance is used to generate earned endowments in dictator games.
 Some works based on data from Germany suggest that the cognitive ability of East Germans lagged behind that of West Germans before German unification. The average intelligence quotient (IQ) was 101 among West German military recruits and 95 among East German recruits; the latter gained 0.66 IQ points per year after unification \citep{roivainen2012economic}.
In a similar vein, average Raven test scores were 70 in West Germany and 50 in East Germany \citep{brouwers2009variation}.

\begin{table}[htbp]
\caption{\label{Tab_Raven} Summary Statistics for Raven Test Z-Scores}\centering\medskip
\begin{tabular}{lcccc} \hline \hline
 & mean  & SD  & min  & max  \\  \hline 
NK &     -.771 &       .79 &     -1.96 &      1.13 \\  
SK &      .724 &      .572 &     -1.52 &      1.57 \\  
KC &      .104 &      .904 &     -1.82 &      1.43 \\  
Total & 0 &         1 &     -1.96 &      1.57 \\  
\hline \end{tabular}
\end{table}

 In Figure \ref{Fig_Raven},
the estimated age profiles show a persistent SK-NK gap across ages, with KC participants generally between the two groups.
The steeper profile for KC participants may reflect cohort differences associated with China's rapid economic development since the 1990s. Potential contributors include improvements in nutrition, educational access and quality, and returns to education \citep{Wu:10, Jianjun:12, Heckman:02}.

\subsection{Experimental Results}

Table \ref{Tab_experiment_results_01} gives central empirical results at the individual level.  
Panel A of Table \ref{Tab_experiment_results_01} summarizes individual performance by compensation scheme. Subjects performed the
real-effort task twice under two alternative compensation schemes. We
randomized the order of the compensation schemes by session; therefore, about 50\% of subjects
of each country of origin performed under the piece rate scheme first
and under the tournament scheme later (54\% for NK, 50\% for KC and 49\% for
SK).
Under the piece rate scheme, on average, the NK subjects scored about 11 out of 20, which was 2 points below the average scores of SK and KC.
Under the tournament scheme, the NK subjects performed slightly better; however, the SK subjects scored above 13, keeping the gap between SK and NK at about 2 points.   
Under both schemes, on average, the SK subjects performed best, while the NK
subjects performed worst. The KC subjects are between SK and NK. The difference
between NK and SK is statistically significant at the 1\% significance
level. The performance gap is about 18\% in both schemes. There is no
significant gap between SK and KC under the piece rate scheme (less than
0.5\%); however, the gap becomes marginally significant (7\%) under the tournament scheme.

\begin{table}[htbp]
\caption{\label{Tab_experiment_results_01} Empirical Results at the Individual Level}\centering\medskip
\begin{tabular}{lccccc} \hline \hline
 & NK  & SK  & KC  & p-value-NK  & p-value-KC  \\  \hline 
\multicolumn{6}{l}{A. Real effort task performance} \\  \\
Piece rate first &      .539 &      .487 &        .5 &      .307 &      .852 \\  
Score at piece rate &      10.8 &      12.9 &      12.9 &  2.16e-08 &      .905 \\  
Score at tournament &      11.4 &      13.6 &      12.7 &  9.37e-10 &     .0507 \\  
\hline
\multicolumn{6}{l}{B. Compensation scheme choice} \\ \\
Bonus &      4.91 &      4.74 &      5.13 &      .616 &      .358 \\  
Choice of tournament &       .45 &      .648 &      .625 &  .0000898 &      .735 \\  
Score at the chosen scheme &      12.4 &      15.2 &        14 &  3.75e-13 &     .0176 \\  
\hline
\multicolumn{6}{l}{C. Belief elicitation} \\ \\
Subjective winning probability &      .601 &      .781 &      .749 &  6.41e-10 &      .356 \\  
Subjective$-$empirical prob. gap  &     -.119 &    -.0449 &    -.0844 &     .0116 &      .264 \\  
\hline
\multicolumn{6}{l}{D. Lottery choice} \\ \\
Number of safe lottery choices &      4.62 &      5.51 &      4.74 &   .000498 &     .0291 \\  
Inconsistent lottery choices &      .429 &      .104 &      .236 &  7.06e-14 &     .0161 \\  
\hline
\multicolumn{6}{l}{E. Cognitive ability} \\ \\
Raven test score &     -.771 &      .724 &      .104 &  7.25e-70 &  8.53e-08 \\  
\hline  \end{tabular}
\medskip \\
\parbox{5.5in}{Notes: The table shows the mean  of each variable for  North Korean refugees (NK),
native-born South Koreans (SK) and Korean-Chinese immigrants (KC) separately. The p-value-NK and p-value-KC are p-values for testing the significance of NK and KC indicators, respectively, in regressing each variable on group indicators.
The omitted group is SK in the regression.}
\end{table}

Panel B of Table \ref{Tab_experiment_results_01} presents the compensation scheme choice results and task
performance outcomes under the selected scheme. As explained above, subjects
were randomly given some bonus points before choosing a compensation scheme.
The bonus point is a random integer between 0 and 10. Thus, the average is 5
for all three groups, and there is no statistical difference.
The most important finding in Panel B is that the NK subjects were less likely to
choose the tournament scheme than the SK or KC subjects. 45\% of the NK
subjects selected the tournament scheme, whereas 65\% of SK subjects and 63\%
of KC subjects chose the tournament scheme. The difference between SK and
NK or between NK and KC is significantly different from zero at the 1\%
significance level. These choice rates show that NK participants entered the tournament less often than SK or KC participants.
Panel B also shows how well subjects performed in their selected compensation
scheme. Subjects performed better in the selected scheme than in Panel A. This difference may reflect learning from repeating the same type of task and self-selection into the preferred compensation scheme.
As in Panel A, the NK subjects performed worst, while the SK and KC subjects scored 15 and 14 on average, respectively.



Panel C of Table \ref{Tab_experiment_results_01} reports elicited subjective winning probability across the three groups.
The NK subjects, on average,  assessed their chance of winning with a bonus point at 60\%; whereas the average winning probability was 
78\% and 75\% for SK and KC, respectively. The average gap between the subjective and ex post empirical winning probability was negative for all the groups; NK showed the most pessimistic self-evaluation of the likelihood of winning. 
NK participants both entered the tournament less often and reported lower probabilities of winning than the other groups.

Panel D of Table \ref{Tab_experiment_results_01} summarizes the elicited risk preferences. Out of ten choices between paired lotteries, all three groups selected about five safe lotteries, with a somewhat higher average for SK. The more noticeable difference is that inconsistent lottery choices due to multiple switching
 is much more frequent for NK than SK and KC: 82 out of 191 NK subjects made an inconsistent lottery choice; only 57\% of NK were consistent, whereas only 20 out of 193 SK subjects made an inconsistent lottery choice. Like many other variables, the result for KC is between those for SK and NK. Panel E reports the Raven test scores alongside these outcomes for comparison.

\subsection{Determinants for Selection into Tournament}

\begin{table}[htbp]
\caption{\label{Tab_experiment_results_02} Determinants for Selection into Tournament}\centering\medskip
\begin{tabular}{lccccc} \hline\hline
 & (1) & (2) & (3) & (4) & (5) \\
\hline
 &  &  &  &  & \\
NK & -.171 & -.156 & -.11 & .0893 & .122 \\
 & (.0384) & (.0346) & (.0441) & (.0561) & (.0647) \\
KC & -.0513 & -.0525 & -.0285 & .0644 & .107 \\
 & (.0901) & (.0823) & (.0777) & (.07) & (.0799) \\
Age & -.0139 & -.0131 & -.0112 & -.00659 & -.00582 \\
 & (.0032) & (.00325) & (.00329) & (.00302) & (.0027) \\
Female & -.109 & -.0862 & -.0925 & -.0757 & -.0971 \\
 & (.054) & (.0551) & (.0543) & (.0504) & (.0555) \\
Bonus & .0321 & .0308 & .0306 & .0327 & .0348 \\
 & (.00529) & (.00506) & (.00491) & (.00503) & (.00556) \\
Number of safe lottery choices &  & -.0219 & -.0216 & -.0198 & -.0198 \\
 &  & (.00859) & (.00824) & (.00794) & (.00706) \\
Inconsistent lottery choices &  & -.107 & -.103 & -.0532 & -.0403 \\
 &  & (.0504) & (.0547) & (.0561) & (.06) \\
Pre-choice tournament score &  &  & .0231 & .0161 & .0135 \\
 &  &  & (.00589) & (.00606) & (.00681) \\
Raven test score &  &  &  & .162 & .16 \\
 &  &  &  & (.0359) & (.0351) \\
   &  &  &  &  &  \\
Further sociodemographic variables &  No &  No &  No &  No & Yes \\ 
  &  &  &  &  &  \\
Observations & 456 & 456 & 456 & 456 & 423 \\
 R-squared & .165 & .185 & .211 & .25 & .29 \\ \hline
\end{tabular}
\medskip \\
\parbox{5.5in}{Notes: 
Linear probability models. The dependent variable is an indicator of whether the subject selects the tournament scheme.  
Further sociodemographic variables include 
post-secondary education, 
marital status, subjective health status, 
religious affiliation,  
number of household members,
labor force status,  
stock market participation, 
credit card holding, 
online shopping, 
log monthly household income, 
log monthly household expenditure,
and 
log household wealth. 
Robust standard errors, clustered by session, are presented in parentheses.}
\end{table}

We next examine how the estimated group differences in tournament entry change as controls are added.
As mentioned earlier and shown in Table \ref{Tab_data_01}, the three Korean groups,
particularly, SK and NK, differed along important observed and unobserved
dimensions. To address this concern, in the regression analysis below, we
controlled for some demographic and socioeconomic characteristics. Furthermore,
we elicited or measured additional characteristics relevant to the compensation-scheme choice, such as risk aversion and ability. We estimated conditional differences in tournament entry after
controlling for these characteristics.

Table \ref{Tab_experiment_results_02} presents the regression results. In column (1), we controlled
for age, a female indicator, and bonus points only. We then added control variables sequentially to check the
robustness of the inter-group differences. Column (1) shows that the NK subjects
were about 17 percentage points less likely to select the tournament scheme
than the SK subjects while there was no significant difference between the SK and KC
subjects. In column (2), we controlled for two lottery-choice measures: the number of safe choices and whether participants made inconsistent choices (more than one switching point). The NK-SK difference
declined slightly to 15.6 percentage points but remained statistically
significant. This result holds in column (3) where we controlled for the level of pre-choice task performance.

The NK-SK gap became statistically insignificant in column (4)
once we controlled for the Raven test score, a measure of general cognitive ability.
The NK coefficient changed sign but remained statistically insignificant. The magnitude of the negative age coefficient fell by approximately half and also became statistically insignificant when the Raven score was included.
 The KC-SK coefficients were statistically insignificant across all specifications in Table \ref{Tab_experiment_results_02}.
The results in column (4) are similar to those in column (5), which includes further sociodemographic variables.

In the post-experiment survey, we asked a question about preference for
competition. The exact question was ``What do you think about competing with
others in a usual day?'' The response was from 1 (hate it very much) to 10
(like it very much).\footnote{A general question on competition is adopted here to avoid any potential bias against NK or KC. In a different context, 
a general risk-taking question was found to be a broadly applicable predictor of risky behavior \citep{Dohmen:11}.}
 The survey results show that the average was highest for
NK subjects (6.6) and lowest for SK subjects (5.5). KC subjects' average was
around the middle, 6.2. NK participants therefore reported more favorable stated attitudes toward everyday competition than SK participants. This difference remained after controlling for individual characteristics, including general cognitive ability. The stated attitudes do not mirror the observed NK-SK difference in incentivized tournament entry.
KC participants also reported more favorable stated attitudes toward competition than SK participants.

\subsection{Determinants for Subjective Winning Probability}\label{sec:winprob}

\begin{table}[htbp]
\caption{\label{Tab_experiment_results_03} Determinants for Subjective Winning Probability}\centering\medskip
\begin{tabular}{lccccc} \hline\hline
 & (1) & (2) & (3) & (4) & (5) \\
\hline
 &  &  &  &  & \\
NK & -.167 & -.154 & -.106 & -.00794 & -.0236 \\
 & (.0244) & (.0264) & (.0273) & (.0279) & (.0279) \\
KC & -.0515 & -.0469 & -.0217 & .024 & .0251 \\
 & (.0266) & (.0238) & (.0215) & (.0261) & (.0255) \\
Age & -.00768 & -.00737 & -.00537 & -.00311 & -.00291 \\
 & (.00152) & (.00157) & (.00155) & (.00134) & (.00192) \\
Female & -.0631 & -.057 & -.0636 & -.0554 & -.0608 \\
 & (.0178) & (.0183) & (.018) & (.0162) & (.0165) \\
Bonus & .0255 & .0255 & .0252 & .0263 & .027 \\
 & (.00397) & (.00396) & (.00327) & (.00351) & (.00381) \\
Number of safe lottery choices &  & -.00273 & -.0024 & -.00152 & -.00157 \\
 &  & (.00422) & (.00387) & (.0044) & (.00452) \\
Inconsistent lottery choices &  & -.0477 & -.0439 & -.0192 & -.0157 \\
 &  & (.0354) & (.0349) & (.0334) & (.0378) \\
Pre-choice tournament score &  &  & .0244 & .0209 & .0209 \\
 &  &  & (.00263) & (.00256) & (.00278) \\
Raven test score &  &  &  & .0794 & .0763 \\
 &  &  &  & (.0186) & (.0184) \\
   &  &  &  &  &  \\
Further sociodemographic variables &  No &  No &  No &  No & Yes \\ 
  &  &  &  &  &  \\
Observations & 456 & 456 & 456 & 456 & 423 \\
 R-squared & .248 & .253 & .339 & .367 & .366 \\ \hline
\end{tabular}
\medskip \\
\parbox{5.5in}{Notes: 
Linear regressions.
Further sociodemographic variables are the same as those in Table \ref{Tab_experiment_results_02}. 
Robust standard errors, clustered by session, are presented in parentheses.}
\end{table}

Panel C of Table \ref{Tab_experiment_results_01} shows that NK participants reported a subjective probability of winning 18 percentage points lower than that reported by SK participants. The corresponding KC-SK difference is not statistically significant.
Two factors may contribute to this pattern. First, prior tournament performance provides information about likely performance in a subsequent tournament. Participants who performed poorly may therefore report lower winning probabilities and choose the piece-rate scheme more often.\footnote{Participants nevertheless had limited information about the distribution of ability among others in their session, making precise assessments of winning probabilities difficult.} Because NK participants performed below average, performance differences may account for part of their lower reported winning probabilities and tournament entry. Second, subjective beliefs may reflect self-confidence as well as performance information. Self-confidence varies substantially across individuals \citep{barber2001boys}; lower confidence could reduce reported winning probabilities conditional on performance. Aversion to competition could separately reduce tournament entry conditional on beliefs. This subsection reports reduced-form associations between individual characteristics and subjective winning probability; Section \ref{sec:structural:est} considers these components jointly within the choice model.

Table \ref{Tab_experiment_results_03} examines which individual characteristics are associated with the NK-SK difference in subjective winning probability.
Column (1) includes age, gender, and bonus points. NK participants report a subjective winning probability 17 percentage points below that of SK participants. The corresponding KC-SK difference is 5.2 percentage points and marginally significant. Adding two measures of risk aversion in column (2) reduces the NK-SK difference to 15.4 percentage points. Column (3) adds pre-choice task performance, which is significantly associated with subjective winning probability and reduces the difference to 10.6 percentage points.\footnote{We also considered two indicators from the post-experiment survey: whether participants expected their own group to perform best or worst among the NK, SK, and KC groups. These indicators proxy for group-level confidence. Participants who expected their own group to perform worst were less likely to enter the tournament, but the indicators did not account for the NK-SK difference.} The remaining difference is statistically significant, suggesting that prior performance alone does not account for the gap in subjective beliefs.

In column (4), we controlled for the Raven test score. The NK-SK difference in subjective winning probability became statistically
insignificant and virtually zero in magnitude. Thus, the Raven score statistically accounts for the observed difference in subjective winning probability, as it does for the NK-SK difference in
tournament entry.
As before, the inclusion of further sociodemographic variables in column (5) gives results similar to those in column (4).

\subsection{Associations with Measured Experiences in North Korea and South Korea}

This subsection examines associations between tournament entry and NK participants' measured experiences in North Korea and South Korea. The post-experiment survey collected detailed information about their economic activities and experiences in North Korea. Despite the strong constraints imposed by North Korea's social and political system, participants varied in their economic and social experiences, including their exposure to markets and their positions within its class structure.

\begin{table}[htbp]
\caption{\label{table_nk} Descriptive Statistics for NK Specific Variables}\centering\medskip
\begin{tabular}{lcccc} \hline \hline
 & mean  & SD  & min  & max  \\  \hline 
Border provinces &      .791 &      .408 &         0 &         1 \\  
Military service in NK &      .147 &      .355 &         0 &         1 \\  
Communist party member in NK &      .131 &      .338 &         0 &         1 \\  
Secondary job in NK &      .408 &      .493 &         0 &         1 \\  
Years of secondary job in NK &       1.4 &      3.12 &         0 &        30 \\  
Years in SK &      7.12 &      3.78 &       .75 &        15 \\  
Education in SK &       .45 &      .499 &         0 &         1 \\  
Monthly household income &       1.5 &      1.48 &        .1 &        12 \\  
\hline \end{tabular}
\end{table}

Table \ref{table_nk} summarizes the NK-specific variables. Almost 80\% of NK participants were born in border provinces, 15\% served in the military, 13\% were Communist Party members, and 40\% had held a secondary job. Including participants with no such experience, the average duration of secondary employment was 1.4 years. Secondary jobs comprise income-generating activities conducted mainly through markets, including trade, production of consumer goods, smuggling, repair work, private services, livestock feeding, and cultivation of private plots \citep{Kim2017}. Participants had spent an average of about seven years in South Korea, 45\% had received some education there, and average monthly household income was 1.5 million KRW.

\begin{table}[htbp]
\caption{\label{Tab_experiment_results_02_NK} Determinants for Selection into Tournament Using NK Specific Variables}\centering\medskip
\begin{tabular}{lcccc} \hline\hline
 & (1) & (2) & (3) & (4) \\ \hline
 &  &  &  &  \\
NK & .0893 & .0537 & .0188 & -.071 \\
 & (.0561) & (.106) & (.0629) & (.104) \\
KC & .0644 & .0653 & .0538 & .0541 \\
 & (.07) & (.0688) & (.0728) & (.0708) \\
Age & -.00659 & -.00745 & -.00722 & -.00835 \\
 & (.00302) & (.00266) & (.00302) & (.00277) \\
Female & -.0757 & -.0693 & -.0583 & -.0428 \\
 & (.0504) & (.0494) & (.0475) & (.0493) \\
Bonus & .0327 & .0327 & .0359 & .0361 \\
 & (.00503) & (.0054) & (.0046) & (.00517) \\
Number of safe lottery choices & -.0198 & -.0193 & -.0208 & -.0205 \\
 & (.00794) & (.0082) & (.00843) & (.00888) \\
Inconsistent lottery choices & -.0532 & -.0588 & -.0545 & -.0616 \\
 & (.0561) & (.055) & (.0558) & (.056) \\
Pre-choice tournament score & .0161 & .0166 & .0142 & .0149 \\
 & (.00606) & (.0061) & (.00634) & (.0065) \\
Raven test score & .162 & .162 & .152 & .151 \\
 & (.0359) & (.0373) & (.033) & (.0342) \\
NK $\times$ (Border provinces) &  & .0562 &  & .0734 \\
 &  & (.0693) &  & (.0695) \\
NK $\times$ (Military service in NK) &  & -.105 &  & -.0739 \\
 &  & (.141) &  & (.147) \\
NK $\times$ (Communist party member in NK) &  & .178 &  & .211 \\
 &  & (.156) &  & (.165) \\
NK $\times$ (Secondary job in NK) &  & -.08 &  & -.0692 \\
 &  & (.0735) &  & (.0648) \\
NK $\times$ (Years of secondary job in NK) &  & .0162 &  & .0235 \\
 &  & (.0131) &  & (.0116) \\
NK $\times$ (Years in SK - 7) &  &  & .0258 & .0278 \\
 &  &  & (.00725) & (.00786) \\
NK $\times$ (Education in SK) &  &  & .0762 & .0992 \\
 &  &  & (.0533) & (.0583) \\
NK $\times$ (log(household income in SK) - log(1.5)) &  &  & -.0544 & -.0796 \\
 &  &  & (.0491) & (.0488) \\
 &  &  &  &  \\
Observations & 456 & 456 & 456 & 456 \\
 R-squared & .25 & .258 & .271 & .283 \\ \hline
\end{tabular}
\medskip \\
\parbox{5.5in}{Notes: 
Robust standard errors, clustered by session, are presented in parentheses.}
\end{table}

Table \ref{Tab_experiment_results_02_NK} presents specifications based on
Table \ref{Tab_experiment_results_02}. We begin with the baseline specification in column
(4) in Table \ref{Tab_experiment_results_02} and add interactions between the NK indicator variable and NK-specific variables. In column (2) of Table \ref{Tab_experiment_results_02_NK}, none of the estimated interaction coefficients for the measured experiences in North Korea is individually statistically significant, and a joint test fails to reject that they are all zero. These variables include Communist Party membership, a proxy for socioeconomic status in North Korea, and indicators of potential exposure to markets through residence in a border province or a secondary job. The estimates leave considerable uncertainty about how these experiences relate to tournament entry.
In column (3), we include NK refugees' experiences in South Korea:
the duration of residence in SK, education in SK and monthly household income. Longer residence in South Korea is associated with a higher probability of tournament entry. This pattern is consistent with assimilation but could also reflect differences across arrival cohorts. For subjective winning probability, none of the measured experiences in North Korea or South Korea has a statistically significant association.

\subsection{Possible Sources of Raven-Score Differences}

The NK-SK coefficient in the tournament-entry regressions becomes statistically insignificant after the Raven score is included. This result raises the question of what underlies the Raven-score gap. Raven scores may reflect education, childhood health and nutrition, socioeconomic conditions, test familiarity, and other factors. Education is one possible contributor. \citet{Kim:Lee:18} compare years of schooling and Raven scores across countries and find that North Koreans have lower cognitive skills than populations in comparable less-developed countries despite longer schooling. They point to the content and quality of education in North Korea, including an emphasis on ideological instruction and the mobilization of students for collective labor \citep{UNK:2014}.

	Childhood nutrition is another possible contributor. \citet{Kim:Lee:18} find that observed demographic and socioeconomic characteristics do not account for the variation in cognitive ability among North Korean refugees and point to macro-level conditions such as economic development and food shortages.
	
Related research links early childhood conditions to human-capital development \citep{Currie} and documents anthropometric differences between North and South Koreans. \citet{Schwekendiek} find that North Korean preschool children are about 6-7 cm shorter and 3 kg lighter than South Korean children, and that North Korean refugee children and adolescents are about 3-4 cm shorter and 1 kg lighter than their South Korean counterparts. \citet{Pak} finds that time spent in South Korea or a third country after leaving North Korea is associated with partial catch-up. Together, these studies make education and childhood conditions plausible contributors to Raven-score differences, while leaving their relative importance unresolved in our sample.

Against this background, we next examine how the Raven score and the observed difference in tournament entry relate to the components of choice within the model.

\section{A Choice Model of Tournament Entry with Probability Weighting}\label{sec:structural:est}

Following recent discussions about variation in willingness to compete \citep{gillen2019experimenting, Veldhuizen}, we use a simple model of choice between the piece-rate scheme ($PS$) and the tournament scheme ($TS$) to organize four components of tournament entry: expected performance, subjective beliefs about winning, risk preferences, and a residual competitive component not accounted for by the other three. The resulting decomposition provides a model-based accounting of how the Raven score is associated with tournament entry.

Assume that participant $i$ predicts tournament performance after observing the random bonus as
\[
x_{i} = p_{i} + b_{i},
\]%
where
$x_{i}$ is the predicted tournament score, $p_{i}$ is expected performance,
and $b_{i}$ is the random bonus.

Suppose that participant $i$'s utility from choosing the piece-rate scheme is
\begin{align}\label{util-PS}
U_i \left( PS \right) =u_i \left( x_{i} \right) -c\left(
x_{i}\right), 
\end{align}
where the utility function over money is
$u_i \left( x \right) =x^{\alpha_i }
$
and 
$c\left( x_{i} \right) $ denotes the disutility of making
an effort to solve $x_{i}$.\footnote{It is implicitly assumed that $x_i > 0$ for every $i$.}
Here, $u_i \left( \cdot \right)$ allows for individual heterogeneity via  
$\alpha_i$, which represents participant $i$'s risk preference, whereas
$c \left( \cdot \right)$ is assumed to be homogeneous across subjects. 
This specification uses the observed lottery choices to allow for heterogeneous risk preferences, while effort costs remain unobserved and are treated as homogeneous.

We consider the following specification for the expected utility
of choosing the tournament scheme:
\begin{align}\label{util-TS}
U_i \left( TS \right) =u_i\left( 2x_{i}\right) \times w_i \left( \Pr
\left\{ x_{i}>p_{j}\right\} +\frac{\Pr \left\{ x_{i}=p_{j}\right\} }{2}%
\right) -c\left( x_{i}\right) +\phi_i, 
\end{align}
where 
$u_i \left( \cdot \right)$ and $c \left( \cdot \right)$ are  the same as in the piece-rate scheme,
$p_{j}$ is the expected performance of opponent $j$, 
$w_i\left( \cdot \right) $ is a probability weighting function,
and $\phi_i$ is the residual competitive component of tournament utility: within the maintained model, it is the component not represented by performance, probability weighting, risk preferences, or common effort costs.
As with $u_i \left( \cdot \right)$, we allow individual heterogeneity in both
 $w_i\left( \cdot \right) $ and $\phi_i$.
In evaluating the probability in \eqref{util-TS},  $p_{j}$ is compared to $x_{i}$ because the random bonus is applied only to subject $i$, and not to opponent $j$, when determining a winner in the tournament.
Finally,  subject $i$'s choice, say $d_i$, is determined by
\[
d_i = TS \; \text{ iff } \; U_i \left( TS \right)  > U_i \left( PS \right),
\]
equivalently
\begin{align}\label{struc:model:spec}
d_i = TS \; \text{ iff } \;
u_i\left( 2x_{i}\right) \times w_i \left( \Pr
\left\{ x_{i}>p_{j}\right\} +\frac{\Pr \left\{ x_{i}=p_{j}\right\} }{2}%
\right) -u_i \left( x_{i} \right)  +\phi_i > 0.
\end{align}
We next describe the components of the choice model and their estimation.

\subsection{Expected performance}

\begin{figure}[h!tb]
\caption{\label{Fig_AvgP} Average Performance and Random Bonus by Group}\centering
\medskip
 \includegraphics[scale=0.4]{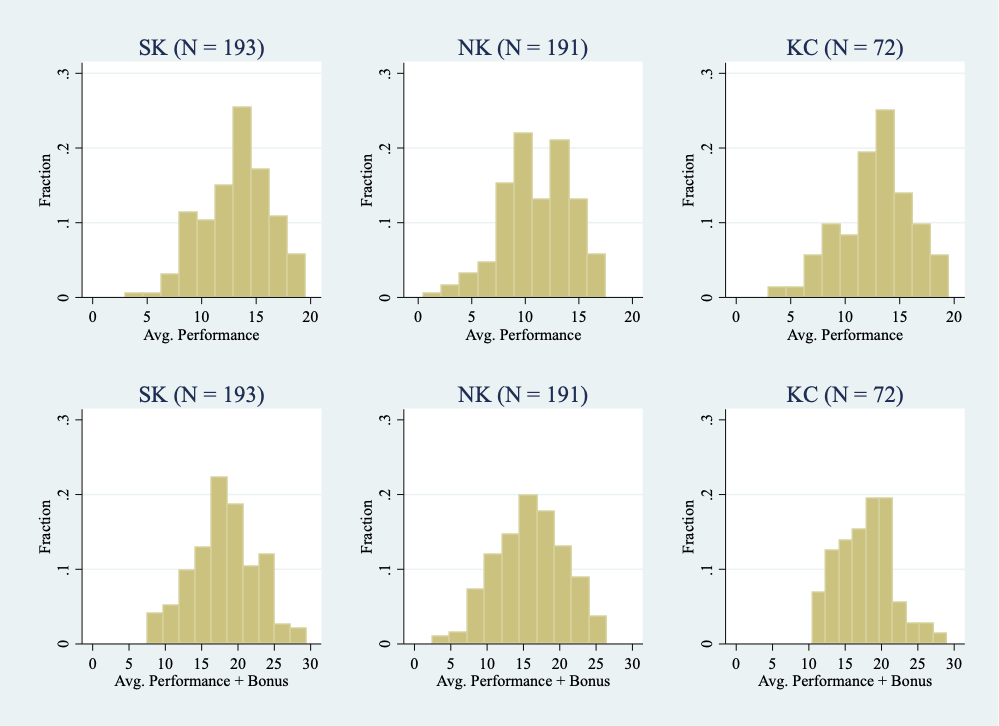} 
\parbox{5.5in}{Notes: 
The average performance refers to the average number of correct answers during the first two individual real-effort tasks. The top panel shows histograms of the average performance by group, while the bottom panel depicts
histograms of the average performance plus the random bonus by group.}
\end{figure}

We measure $p_i$ using the average number of correct answers during the first two individual real-effort tasks. Figure \ref{Fig_AvgP} shows that SK and KC participants performed noticeably better than NK participants and that adding the random bonus substantially diluted the group differences.

\begin{table}[h!tb]
\caption{\label{Tab_structural_results_01} Average Performance before Selecting an Incentive Scheme}\centering\medskip

\begin{tabular}{lccc} \hline\hline
 & (1) & (2) & (3) \\
\hline
 &  &  &   \\
Piece rate first & -.498 & -.234 & -.248 \\
 & (.318) & (.312) & (.305) \\
NK & -2.17 & -1.95 & -.271 \\
 & (.313) & (.293) & (.316) \\
KC & -.499 & -.6 & .185 \\
 & (.688) & (.63) & (.546) \\
Age &  & -.0847 & -.0442 \\
 &  & (.0178) & (.0147) \\
Female &  & .166 & .315 \\
 &  & (.24) & (.254) \\
Raven test score &  &  & 1.19 \\
 &  &  & (.137) \\
Constant & 13.5 & 16.2 & 13.8 \\
 & (.228) & (.624) & (.486) \\
 &  &  &  \\
Observations & 456 & 456 & 456 \\
 R-squared & .1 & .158 & .21 \\ \hline
\end{tabular}
\medskip \\
\parbox{5.5in}{Notes: 
The dependent variable is the average number of correct answers in the first two individual real-effort tasks. Robust standard errors, clustered by session, are presented in parentheses.}
\end{table}

Table \ref{Tab_structural_results_01} reports regressions for average performance. Column (1) includes group indicators and an indicator for whether the piece-rate task came first. Column (2) adds age and a female indicator, and column (3) adds the Raven score.
NK participants average about two fewer correct answers in columns (1) and (2). After the Raven score is included, the NK coefficient falls to $-0.271$ and is not statistically significant.

\subsection{Probability weighting}

A central component of \eqref{util-TS} is participant $i$'s probability-weighted winning probability:
\begin{align}\label{def:winning-prob:ideal}
w_i \left( \Pr
\left\{ x_{i}>p_{j}\right\} +\frac{\Pr \left\{ x_{i}=p_{j}\right\} }{2}%
\right).
\end{align}
To operationalize probability weighting,  we build on \citet{goldstein1987expression}'s functional form: 
\begin{align}\label{Goldstein-Einhorn}
w_i\left( p\right) =\frac{\delta_i p^{\gamma }}{\delta_i p^{\gamma }+\left(
1-p\right) ^{\gamma }}
\end{align}
for $\delta_i \geq 0$ and $\gamma \geq 0$. Here, $\delta_i $ captures the degree of
optimism and $\gamma $ the parameter of likelihood insensitivity. 
We allow $\delta_i$ to vary across participants and hold $\gamma$ constant across participants.
When $\delta_i = \gamma = 1$ for each $i$, the model reduces to the expected utility without probability weighting. 

Let $\mathrm{SWP}_i$ denote participant $i$'s reported winning probability elicited using the method of \citet{Hossain:Okui:13}.
We use $\mathrm{SWP}_i$ as a proxy for the probability-weighted winning probability in \eqref{def:winning-prob:ideal}.
The corresponding winning probability based on the empirical performance distribution is
\[
 \Pr
\left\{ x_{i}>p_{j}\right\} +\frac{\Pr \left\{ x_{i}=p_{j}\right\} }{2},
\]
which can be estimated for each $i$ using our experimental data.  Specifically, define
\[
\mathrm{EWP}_i
=
\frac{1}{n} \sum_{j = 1}^n 
\left\{ 
\mathbb{I} ( p_j < p_i + b_i ) 
+ \frac{1}{2} \mathbb{I} ( p_j = p_i + b_i )
\right\},
\]
 where  $\mathbb{I}(\cdot)$ is the usual indicator function and $n$ is the sample size.
 Thus, $\mathrm{EWP}_i$ is the empirical winning probability constructed from the average performance of all participants.
 The probability-weighting function in \eqref{Goldstein-Einhorn} suggests the following regression model:
 \begin{align}\label{swp-ewp-model}
 \log \left( \frac{\mathrm{SWP}_i}{1-\mathrm{SWP}_i} \right)
 = \log \delta_i + \gamma \log \left( \frac{\mathrm{EWP}_i}{1-\mathrm{EWP}_i} \right).
 \end{align}
   We decompose $\log \delta_i$ into observed heterogeneity ($z_i'\theta$) and an idiosyncratic component ($u_i$), where $z_i$ includes the Raven score and other covariates and $\theta$ is the corresponding parameter vector. \citet{CKLL2022} examine how cognitive ability relates to both optimism and likelihood insensitivity. In the present specification, the Raven score enters the optimism parameter $\delta_i$, while the likelihood-insensitivity parameter $\gamma$ is held constant across participants. For some participants, $\mathrm{SWP}_i$ or $\mathrm{EWP}_i$ equals zero or one.
 To have well-defined log odds, we truncate   $\mathrm{SWP}_i$ and $\mathrm{EWP}_i$ to be always between 0.01 and 0.99.
 The unknown parameters $\gamma$ and $\theta$ are estimated via median regression to mitigate the effect of truncation.

\begin{table}[h!tb]
\caption{\label{Tab_structural_results_02} Determinants of Probability Weighting}\centering\medskip

\begin{tabular}{lccc} \hline\hline
 & (1) & (2) & (3) \\ \hline
 &  &  &  \\
 Log odds of $\mathrm{EWP}_i$ & .565 & .501 & .543 \\
 & (.0835) & (.0517) & (.0561) \\ \vspace*{0.5ex}
&[.445,.717]&[.416,.591]&[.425,.612] \\
NK & -1.9 & -1.29 & -.463 \\
 & (.528) & (.28) & (.393) \\ \vspace*{0.5ex}
&[-2.4,-.767]&[-1.84,-.958]&[-.979,.249] \\
KC & -1.43 & -1.04 & -.558 \\
 & (.6) & (.393) & (.432) \\ \vspace*{0.5ex}
&[-2.4,-.379]&[-1.74,-.432]&[-1.2,.187] \\ 
Age &  & -.0465 & -.0235 \\
 &  & (.0117) & (.0125) \\ \vspace*{0.5ex}
  & &[-0.0688,-.0297]&[-.0456,-.00426] \\  
Female &  & -1.08 & -.813 \\
 &  & (.249) & (.283) \\ \vspace*{0.5ex}
  & &[-1.38,-0.564]&[-1.22,-.26] \\ 
Raven test score &  &  & .613 \\
 &  &  & (.181) \\
  & &  &[.397,.971] \\  
 &  &  &  \\
 Observations & 456 & 456 & 456 \\ \hline
\end{tabular}
\medskip \\
\parbox{5.5in}{Notes: 
Median regression. 
The dependent variable is the log odds of subjective winning probability ($\mathrm{SWP}_i$) and 
the first covariate is the log odds of empirical winning probability ($\mathrm{EWP}_i$). 
Bootstrap standard errors and 90\% percentile confidence intervals 
 are presented in parentheses and brackets, respectively.}
\end{table}

Table \ref{Tab_structural_results_02} reports the estimates. Because $\mathrm{EWP}_i$ is constructed from the sample, we bootstrap the entire estimation procedure 1,000 times to obtain standard errors and confidence intervals. In column (1), the coefficient on the log odds of empirical winning probability is $0.565$, corresponding to $\gamma$ in \eqref{swp-ewp-model}. The negative NK and KC coefficients indicate lower reported winning probabilities relative to SK participants, conditional on empirical winning probability. Column (2) adds age and a female indicator. When the Raven score is added in column (3), the NK and KC coefficients become smaller and their confidence intervals include zero, while the female and Raven coefficients remain statistically significant. Overall, reported winning beliefs are positively associated with empirical winning probability, and the group coefficients become smaller after Raven score is included. The negative female coefficient remains sizable after controlling for the Raven score.

\subsection{Risk Preferences and the Residual Competitive Component}

We now estimate the choice model in \eqref{struc:model:spec}. We use the ten lottery choices to allow risk preferences to vary across participants.

\begin{figure}[h!tb]
\caption{\label{Fig_Nsafe} Number of Safe Options by Group}\centering
\medskip
 \includegraphics[scale=0.4]{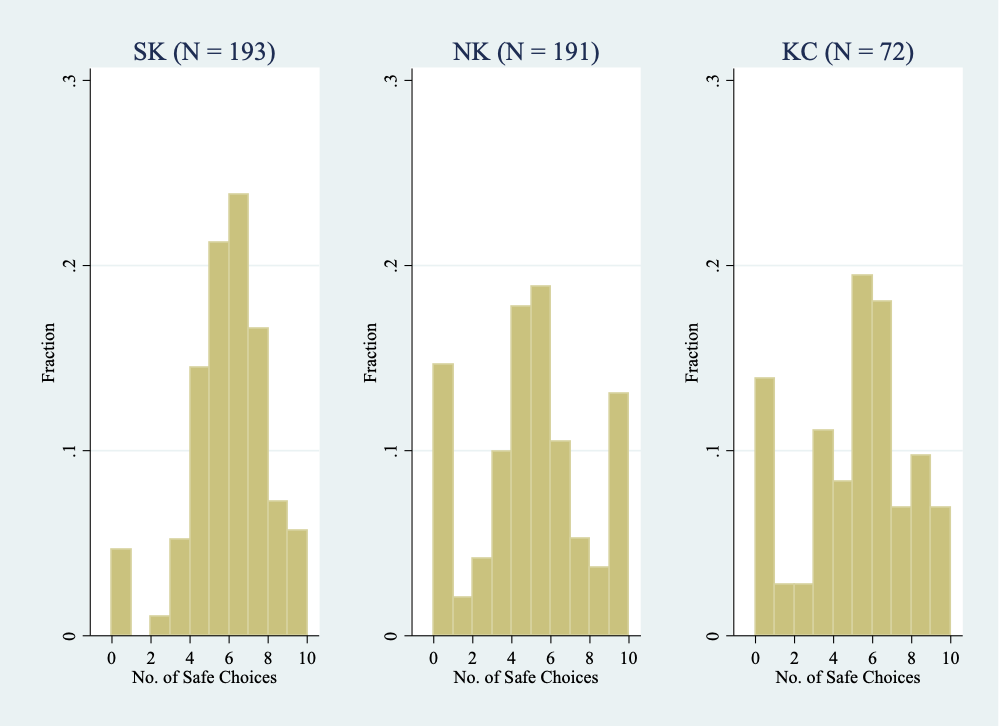} 
\parbox{5.5in}{Notes: 
The histograms show the number of safe options  in the ten paired lottery choices by group.}
\end{figure}

\begin{table}[h!tb]
\caption{\label{Tab_structural_results_03} Patterns in the Number of Safe Options}\centering\medskip

\begin{tabular}{lccc} \hline
 & (1) & (2) & (3) \\ \hline
 &  &  &  \\
NK & -.89 & -.886 & -1.13 \\
 & (.212) & (.183) & (.281) \\
KC & -.772 & -.76 & -.875 \\
 & (.299) & (.303) & (.368) \\
Age &  & .00908 & .0032 \\
 &  & (.0111) & (.0136) \\
Female &  & .57 & .548 \\
 &  & (.285) & (.297) \\
Raven test score &  &  & -.173 \\
 &  &  & (.17) \\
Constant & 5.51 & 4.79 & 5.13 \\
 & (.187) & (.459) & (.717) \\
 &  &  &  \\
Observations & 456 & 456 & 456 \\
 R-squared & .0278 & .0399 & .0418 \\ \hline
\end{tabular}
\medskip \\
\parbox{5.5in}{Notes: 
The dependent variable is the number of safe choices in the ten pairs of lotteries. Robust standard errors, clustered by session, are presented in parentheses.}
\end{table}

Figure \ref{Fig_Nsafe} shows the distribution of safe choices out of the ten lottery pairs. Each group exhibits substantial heterogeneity. Among NK participants, 28 choose no safe option and 19 choose all ten; the corresponding counts are 9 and 1 among SK participants. Ten of the 72 KC participants choose no safe option.
Table \ref{Tab_structural_results_03} reports the corresponding regressions. The negative NK coefficient increases in absolute magnitude after Raven score is included. The coefficients on age, the female indicator, and the Raven score are not statistically significant, and the R-squared in column (3) is 0.04.\footnote{In the literature,
\citet{Dohmen:2010} find that greater risk aversion is associated with lower cognitive ability, whereas
\citet{AHTW:2016} report evidence suggesting that this relation might be spurious.
In Table \ref{Tab_structural_results_03}, the Raven coefficient in the safe-choice regression is not statistically significant. In a separate regression for extreme lottery choice, defined as one when either $\textrm{Safe}_i = 0$ or $\textrm{Safe}_i = 10$, using the same covariates as column (3), the Raven coefficient is $-0.061$ with a standard error of $0.024$. This pattern may reflect decision errors, underlying preferences, or both.}
Recall that the utility function over money is $u_i \left( x \right) =x^{\alpha_i }$. We assign values of $\alpha_i$ according to the number of safe lottery choices:
\begin{align}\label{hetero-risk-aversion}
\begin{split}
u_i(x) 
&= 
    x^2 \times \mathbb{I}(\textrm{Safe}_i \leq 1) + x^{1.5} \times \mathbb{I}(2 \leq \textrm{Safe}_i \leq 3)
+ x \times \mathbb{I}(\textrm{Safe}_i = 4) \\
&+ x^{0.5} \times \mathbb{I}(5 \leq \textrm{Safe}_i \leq 6) 
+ \ln(x) \times \mathbb{I}(7 \leq \textrm{Safe}_i \leq 8)
+ x^{-0.5} \times \mathbb{I}(9 \leq \textrm{Safe}_i \leq 10),
\end{split}
\end{align} 
where $\textrm{Safe}_i$ is the number of safe choices for subject $i$.  
This utility specification follows the broad pattern of the risk-aversion classifications in Table 3 of \citet{Holt:Laury:02}. Allowing utility curvature to vary through \eqref{hetero-risk-aversion} incorporates observed heterogeneity in lottery choices before estimating the residual competitive component.

\begin{table}[h!tb]
\caption{\label{Tab_structural_results_04} Selection into Tournament: Instrumental Variable Estimation}\centering\medskip

\begin{tabular}{lcccccc} \hline
 & (1) & (2) & (3) & (4) & (5) & (6) \\
 & OLS & IV-1 & IV-2 & OLS & IV-1 & IV-2 \\ \hline
 &  &  &  &  &  &  \\
$\textrm{UDiff}_i/1000$ & .298 & 3.16 & 3.54 & .278 & 3.33 & 3.21 \\
 & (.0634) & (1.3) & (.856) & (.0664) & (1.36) & (.86) \\
NK & -.174 & -.257 & -.268 & .0759 & -.0848 & -.0787 \\
 & (.0339) & (.0792) & (.0761) & (.054) & (.109) & (.0917) \\
KC & -.0527 & -.184 & -.202 & .065 & -.109 & -.103 \\
 & (.0902) & (.192) & (.2) & (.0758) & (.196) & (.179) \\
Age & -.0135 & -.00924 & -.00867 & -.00749 & -.00474 & -.00484 \\
 & (.00354) & (.0044) & (.00455) & (.00339) & (.00491) & (.00483) \\
Female & -.107 & .00337 & .018 & -.0856 & .0253 & .0211 \\
 & (.0569) & (.105) & (.1) & (.0511) & (.107) & (.0929) \\
Raven test score &  &  &  & .177 & .125 & .127 \\
 &  &  &  & (.0355) & (.0418) & (.0408) \\
Constant & 1.17 & .785 & .733 & .825 & .513 & .525 \\
 & (.134) & (.229) & (.215) & (.123) & (.255) & (.234) \\ \hline 
 &  &  &  &  &  &  \\ 
First-stage &  &  &  &  &  &  \\
F-statistic &  & 7.48 & 25.31 &  & 7.52  & 27.76 \\
P-value & & .019 & .000 & & .019 & .000 \\
 &  &  &  &  &  &  \\
Observations & 456 & 456 & 456 & 456 & 456 & 456 \\ \hline
\end{tabular}
\medskip \\
\parbox{5.5in}{Notes: 
Linear probability models. The dependent variable is an indicator of whether the subject selects the tournament scheme. $\textrm{UDiff}_i$ is defined in \eqref{def:udiff} and instrumented with the random bonus ($b_i$) in IV-1 or with the empirical winning probability ($\mathrm{EWP}_i$) in IV-2. Robust standard errors, clustered by session, are presented in parentheses.}
\end{table}

In view of \eqref{struc:model:spec}, we approximate the modeled utility difference excluding the residual component by
\begin{align}\label{def:udiff}
\textrm{UDiff}_i = u_i\left( 2x_{i}\right) \times \mathrm{SWP}_i - u_i \left( x_{i} \right). 
\end{align}
We decompose the residual competitive component $\phi_i$ into observed heterogeneity ($z_i'\beta$) and unobserved heterogeneity ($v_i$).
This leads to a regression model of $\mathbb{I}(d_i = TS)$ on $\textrm{UDiff}_i $ and $z_i$.
We instrument $\textrm{UDiff}_i$ with the random bonus ($b_i$) or the empirical winning probability ($\mathrm{EWP}_i$) to address measurement error in $\textrm{UDiff}_i$.

Table \ref{Tab_structural_results_04} reports the estimates. Column (1) presents OLS estimates with group indicators, age, and a female indicator in addition to $\textrm{UDiff}_i$; columns (2) and (3) present the IV estimates. The IV coefficient on $\textrm{UDiff}_i$ is about ten times the OLS coefficient, a pattern consistent with attenuation from measurement error under the IV assumptions.

The NK coefficient of $-0.257$ in column (2) corresponds to a 26-percentage-point lower probability of tournament entry, conditional on the modeled utility difference and other covariates. This coefficient is larger in absolute magnitude than the reduced-form estimates in Table \ref{Tab_experiment_results_02}. The mapping from lottery choices to utility curvature may contribute to this difference: classifications of some NK participants as risk loving increase their modeled utility difference, leaving a larger negative residual NK coefficient. The female coefficients in columns (2) and (3) are small and statistically insignificant. This pattern is consistent with the reduced-form results in Tables \ref{Tab_experiment_results_02} and \ref{Tab_experiment_results_03}, where the female coefficient is negative but insignificant for tournament entry and significantly negative for subjective winning probability. When the Raven score is added in columns (4)-(6), the NK coefficients become small and statistically insignificant, as in the reduced-form tournament-entry regressions.

The choice model organizes the association between the Raven score and tournament entry into expected performance, subjective winning beliefs, and a residual competitive component. Quantitatively, a one-standard-deviation increase in the Raven score is associated with
(i) 1.2 additional correct answers in the real-effort task (column (3) of Table \ref{Tab_structural_results_01}),
(ii) odds of subjective winning that are 1.85 times as large ($1.85 \approx \exp(0.613)$; column (3) of Table \ref{Tab_structural_results_02}),
and (iii) a 13-percentage-point Raven coefficient in the tournament-entry regression conditional on the modeled utility difference and other covariates (columns (5) and (6) of Table \ref{Tab_structural_results_04}). Within the model, the third association is part of the residual competitive component.
In the reduced-form tournament-entry regression, the Raven coefficient is 16 percentage points after controlling for pre-choice tournament performance (columns (4) and (5) of Table \ref{Tab_experiment_results_02}).
Taken together, the estimates associate the Raven score with performance, subjective winning beliefs, and the residual competitive component rather than with a single underlying preference. In the reduced-form regressions, the Raven coefficient is smaller for subjective winning probability than for tournament entry. Within the model, this pattern is consistent with both beliefs and the residual component contributing to tournament entry.

\section{Relation to the Literature}\label{sec:literature}

Our paper contributes most directly to a growing literature examining differences in tournament entry across social groups and cultures. Much of this literature studies gender differences. Women enter tournaments less often than men in many experimental settings
\citep{Gneezy:et:al:03, Gneenzy:Rustichini:04, Niederle:Lise:07, Niederle:Vesterlund:11}.
The opposite pattern appears in a matrilineal society \citep{ECTA971}, suggesting that social environment may shape willingness to compete. \citet{ECOJ12209} also document differences in willingness to compete between fishermen working in cooperative sea-based teams and fishermen working individually at a lake.
Furthermore, communist reforms promoting gender equality may be associated with greater willingness among women to enter tournaments in China
\citep{Booth:2020,Zhang:2020}
and affect women's attitudes toward career success in West Germany \citep{Campa:2019}.

Recent work further distinguishes tournament entry from a separate preference for competition. \citet{Veldhuizen} studies how risk preferences, overconfidence, and competitiveness contribute to gender differences in tournament choices. \citet{Boosey:2026} calibrate task ability, vary whether the distribution of random performance noise is known (risk) or unknown (ambiguity), and use a separate treatment-based measure of preference for competition. Their study focuses on gender differences under risk and ambiguity. By contrast, our tournament combines heterogeneous task performance with a random bonus and examines how cognitive ability, subjective winning beliefs, and group differences relate to tournament entry. The three-group comparison brings this measurement question to populations with sharply different institutional and life experiences.

The setting also relates to research on how the cultural, economic, and political environments in which individuals grow up are associated with preferences, beliefs, and behavior, including trust in financial institutions, stock-market participation, preferences over social policies, willingness to take financial risks, anti-Semitic violence,
political outcomes, and corruption in courts and police. See, for instance,  \citet{Goodbye-Lenin}, \citet{GSZ:04, GSZ:08}, \citet{Osili:Paulson:08}, \citet{Malmendier:Nagel:11}, \citet{VV:2012},
\citet{Grosfeld:15},
and
\citet{Becker:et-al:2016}
 among others.
 \citet{Alesina:Giuliano:2015} review the relationship between culture and institutions.

The paper also relates to research on conditions earlier in the life cycle. Studies of immigration examine the persistence of home-country institutions and social norms after migration and the relationship between cultural distance and economic assimilation
\citep{Friedberg:00, Casey:Dustmann:10, Benabou:11}. 
Related work studies long-run associations between institutions and attitudes within countries. \citet{Cantoni:2017} study how textbook reforms shaped political attitudes in China.
\citet{fuchs2016long} compare East and West Germans after reunification and find that longer exposure to communism is associated with lower investment in human capital and lower labor-market wages.
\citet{Fuchs:2020} analyze
the long-lasting effects of communism in  Eastern Europe.
\citet{Becker:2020} critically examine the selection problems in the context of  
the separation and reunification of Germany.

Finally, the paper contributes to the literature that combines traditional lab experiments with historical contexts.
\citet{Ockenfels:99} and \citet{Brosig:11} examine East-West comparison in the context of
German unification.
\citet{Callen-et-al:14} conduct  experiments on a sample of  Afghanistan civilians to investigate the relationship between violence and economic risk preferences. 
Our previous work \citep{KCLLC, CKLL} compares giving behavior among North Korean refugees and South Korean natives in dictator games.
Using North Korean refugees and South Korean participants recruited in 2016, \citet{Choi:et-al:2024:JDE} find that, among the refugees, stronger implicit bias against South Korea is associated with lower expected profits, lower target profits, and lower realized profits in experimental markets.

\section{Conclusions}\label{sec:conclusion}

This paper compares tournament entry among participants raised in South Korea, North Korea, and China. North Korean refugees enter the tournament about 20 percentage points less often than South Korean participants, and they also enter less often than Korean-Chinese participants. The NK-SK difference remains about 10 percentage points after controlling for risk preferences and task performance measured before the compensation choice. When the Raven score is included, however, the NK-SK coefficient changes sign and becomes statistically insignificant. Conditional on measured cognitive ability and the other controls, the estimates therefore provide no evidence of a difference in tournament entry between the two groups.

The choice model organizes the association between the Raven score and tournament entry into expected performance, subjective winning beliefs, risk preferences, and a residual competitive component. Lower Raven scores are associated with lower expected performance and more pessimistic beliefs. The Raven score is also positively associated with tournament entry after the model accounts for expected performance, beliefs, and risk preferences. Together, these associations show that tournament entry reflects several dimensions of choice.

These comparisons require caution. The three groups differ in observed and unobserved characteristics beyond the institutional environments in which they were raised, and North Korean refugees are a selected and unrepresentative sample of the population living in North Korea. The design therefore does not identify the causal effects of institutional exposure or separately identify the roles of communism, dictatorship, poverty, nutrition, and education. The evidence is best interpreted as a case study of tournament entry across three Korean groups.

More generally, observed tournament-entry gaps across populations may reflect differences in performance, beliefs, risk preferences, probability weighting, cognitive ability, and other factors associated with entry. Examining these components provides a more informative interpretation of such gaps than treating tournament entry as a direct measure of a single competitive preference.

\section*{Ethics statement}

The research reported in this paper was approved on 3 June 2015 by the Institutional Review Board of Seoul National University (IRB No. 1506/002-001).

\section*{Data availability}

The individual-level data are not publicly available because they contain confidential information about participants, including North Korean refugees, for whom disclosure poses heightened privacy and safety risks.

\section*{Funding}

This work was supported in part by the National Research Foundation of Korea Grant funded by the Korean Government (NRF-2013S1A5A2A03044461); the Creative-Pioneering Researchers Program through Seoul National University; the European Research Council (ERC-2014-CoG-646917-ROMIA); and the UK Economic and Social Research Council through research grant ES/P008909/1 to CeMMAP.

The funders had no role in the study design; collection, analysis, or interpretation of the data; preparation of the manuscript; or decision to submit the article for publication.

\section*{Declaration of generative AI and AI-assisted technologies in the manuscript preparation process}

During the preparation of this work, the authors used OpenAI Codex to assist with manuscript organization and language editing. After using this service, the authors reviewed and edited the content as needed and take full responsibility for the content of the article.

\section*{Acknowledgments}

We thank Il Myoung Hwang for helpful comments and Philip Jang for excellent research assistance.

{\singlespacing
\bibliographystyle{economet}
\bibliography{NK-defect}
}

\clearpage
\includepdf[pages=-,pagecommand={}]{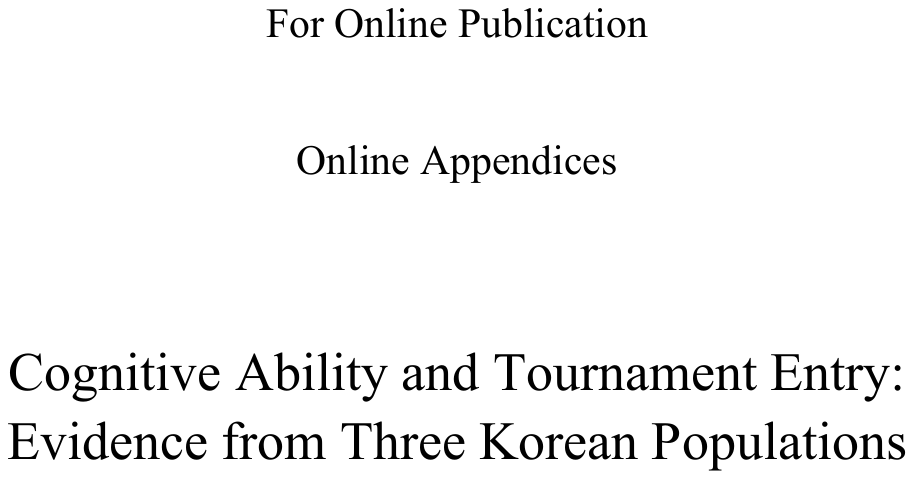}

\end{document}